\documentclass[pre,aps,floats,floatfix,superscriptaddress,nofootinbib,showpacs]{revtex4}

\usepackage[dvips]{graphicx}
\usepackage[activeacute,english]{babel}
\usepackage{eufrak,colordvi,amsmath,epsfig,color}
\newcommand{\be}{\begin{equation}}
\newcommand{\ee}{\end{equation}}

\begin{document}

\title{Fluctuating hydrodynamics 
for driven granular gases}

\author{Pablo Maynar}
\affiliation{F\'{\i}sica Te\'{o}rica, Universidad de Sevilla,
Apartado de Correos 1065, E-41080, Sevilla, Spain}
\author{Mar\'{\i}a Isabel Garc\'{\i}a de Soria}
\affiliation{F\'{\i}sica Te\'{o}rica, Universidad de Sevilla,
Apartado de Correos 1065, E-41080, Sevilla, Spain}
\author{Emmanuel Trizac}
\affiliation{LPTMS (CNRS UMR 8626), Universit\'e Paris-Sud, Orsay Cedex, F-91405, France}
\date{\today }

\begin{abstract}
We study a granular gas heated by a stochastic thermostat in
the dilute limit. Starting from the kinetic equations governing the 
evolution of the
correlation functions, 
a Boltzmann-Langevin equation is constructed. The spectrum
of the corresponding linearized Boltzmann-Fokker-Planck 
operator is analyzed, and the
equation for the fluctuating transverse velocity is derived in the
hydrodynamic limit. The noise term (Langevin force) 
is thus known microscopically and
contains two terms: one coming from the thermostat and the other from the
fluctuating pressure tensor. At variance with the free cooling situation,
the noise is found to be white and its amplitude is evaluated.
\end{abstract}

\maketitle

\section{Introduction}

Typically, a granular system is defined as an ensemble of
macroscopic particles which collide inelastically, i.e. part of the kinetic
energy of the grains is dissipated in a collision. This simple ingredient
gives rise to a very rich phenomenology which is of interest not only from
practical or industrial perspective, but also because of the resulting 
new theoretical 
challenges \cite{d00,g03,bte05,at06}. One of the most widely
employed idealized model for granular fluids is a system of smooth hard
spheres (or disks in two dimensions) whose collisions are characterized by a
constant coefficient of normal restitution \cite{gz93,my96}. For this model,
and considering 
that the particles move freely between collisions, kinetic equations have been
derived: starting from the dynamics of the particles, it is possible to 
derive the corresponding Liouville equation, and the Boltzmann equation results
in the low density limit \cite{bds97,vneb98}. This kinetic equation has been
extensively 
used to address many fundamental questions such as the derivation of the
hydrodynamic equations, with explicit expressions for the transport
coefficients, which have been  derived by the Chapman-Enskog method
\cite{bdks98,sg98} and also via the linearized Boltzmann equation \cite{bdr03}.
Due to the inelasticity of the collisions, the total energy of an isolated
granular system decays 
monotonically in time. In the fast-flow regime, it has been shown numerically
that, for a wide class of initial conditions, the system reaches the so-called
\emph{Homogeneous Cooling State} (HCS), in which all the time dependence of
the one-particle distribution function goes through the granular temperature,
which is defined as the second velocity moment of the distribution
\cite{gs95,h83}. This state has been extensively studied in the literature and
very recently the  
fluctuations of the transverse velocity have been analyzed
\cite{bgm08,bmg09}. It has 
been found 
that the transverse velocity fulfills a Langevin equation but, in contrast to
the elastic case, the noise is
not white and the second moment of the fluctuations is not only controlled by
the viscosity but also depends on a new coefficient. Similar results are
found for the other hydrodynamic equations \cite{betal09}. The study of
fluctuations in the HCS is 
important for the development of a general theory of fluctuations in granular
systems because it defines the reference state from which macroscopic
hydrodynamic equations can be derived \cite{bdks98}. In this sense, the
HCS plays, for inelastic gases, a role similar to the equilibrium state for
molecular gases. 

On the other hand, there are situations in which the grains cannot be
considered to move freely between collisions. If, for example, the grains are
immersed in a medium which acts as a thermostat, the system may reach a
stationary state in which the energy injected by the thermostat is compensated
by the energy dissipated in collisions. Note that, if the grains are Brownian
particles, the interstitial medium injects energy into the granular system, 
but also acts as an energy sink
due to frictional forces. One of the simplest mechanism that can
be considered to thermalize the system is a white noise force 
acting on each grain,
which results in the so-called stochastic thermostat
\cite{vne98,wm96,plmv99,vnetp99,ptvne01,ms00,mss01,gm02,vpbvwt06,etb06,etbb06,faz09}.
One important point is that the
distribution function differs from that of the HCS \cite{vne98}, and that it is 
this distribution which plays the role of the ``reference state''. The
non-equilibrium steady state that the system reaches in the long time limit
exhibits long-range correlations which are in agreement with the predictions of
fluctuating hydrodynamics \cite{vnetp99}. The latter description
was introduced phenomenologically, and is expected 
to be valid in the vicinity of the elastic limit only. 
The objective of this work is to derive
these equations from a more fundamental point of view and without the
restriction of small inelasticity. More precisely, we adapt the formalism 
worked out
in \cite{bmg09} for the free cooling, to the present driven case. 
Starting from a
Boltzmann-Langevin description, we derive a fluctuating equation for the
transverse velocity identifying the noise of this equation. Under certain
hypothesis to be clarified in the text, we obtain that the correlation
function of the noise is well approximated by the one introduced in
Ref. \cite{vnetp99}, where the internal noise contribution 
(excluding the ``external'' noise term directly stemming from the thermostat)
fulfilled a fluctuation-dissipation relation as for conservative
fluids \cite{Landau}.

The remainder of the paper is organized as follows. In Sec. \ref{preliminary}
previous results for a system heated by a stochastic thermostat are presented,
such as the equations for  one-particle distribution function and the
two-particle correlation function. In Sec. \ref{ecboltzmannlangevin} the
Boltzmann-Langevin equation for this system is derived and the properties of
the noise are inferred. The particular case of the transverse velocity field
is analyzed in Sec. \ref{transversevelocity} and finally, the conclusions are
presented in Sec. \ref{conclusions} 

\section{Stochastic thermostat: preliminary results}\label{preliminary} 

The system considered is a dilute gas of $N$ smooth inelastic hard particles
of mass $m$ and diameter $\sigma$. The position and velocity of the $i$th
particle at time $t$ will be denoted by $\mathbf{R}_i(t)$ and
$\mathbf{V}_i(t)$, respectively. The effect of a collision between two
particles $i$ and $j$ is to instantaneously modify their velocities according
to the collision rule
\begin{eqnarray}
\mathbf{V}_i'&=&\mathbf{V}_i-\frac{1+\alpha}{2}
(\hat{\boldsymbol{\sigma}}\cdot\mathbf{V}_{ij})\hat{\boldsymbol{\sigma}},
\nonumber\\
\mathbf{V}_j'&=&\mathbf{V}_j+\frac{1+\alpha}{2}
(\hat{\boldsymbol{\sigma}}\cdot\mathbf{V}_{ij})\hat{\boldsymbol{\sigma}},
\end{eqnarray}
where $\mathbf{V}_{ij}\equiv\mathbf{V}_i-\mathbf{V}_j$ is the relative
velocity, $\hat{\boldsymbol{\sigma}}$ is the unit vector pointing from the
center of particle $j$ to the center of particle $i$ at contact, and $\alpha$
is the coefficient of normal restitution. It is defined in the interval 
$0<\alpha\le 1$ and it will be considered here as a constant, independent of
the relative velocity. Between collisions, the system is heated uniformly by
adding a random velocity to the 
velocity of each particle at equal times. 
The driving is implemented in such a way that the time between random kicks is
small  compared to the mean free time. Then, between collisions, the
velocities of the particles undergo a large number of kicks 
due to the thermostat. In addition, we will assume that the
``jump moments'' of the velocities of the particles verify 
\begin{eqnarray}\label{noise}
B_{ij,\beta\gamma}\equiv\lim_{\Delta t\to 0}
\frac{\langle\Delta V_{i,\beta}\Delta V_{j,\gamma}\rangle_{noise}}{\Delta t}
=\xi_0^2\delta_{ij}\delta_{\beta\gamma}
+\frac{\xi_0^2}{N}(\delta_{ij}-1)\delta_{\beta\gamma},\\
i,j=1,\dots,N \qquad \text{and} \qquad \beta,\gamma=1,\dots,d\nonumber
\end{eqnarray}
where we have introduced 
$\Delta V_{i,\beta}\equiv V_{i,\beta}(t+\Delta t)- V_{i,\beta}(t)$,  
$ V_{i,\beta}(t)$ being the $\beta$ component of the velocity of particle $i$
at time $t$. We have also introduced the strength of the noise, $\xi_0^2$, and 
$\langle\dots\rangle_{noise}$, which denotes average over different
realizations of the noise. The non-diagonal terms (corresponding to $i\ne j$
and $\beta=\gamma$) are necessary in order to conserve the total momentum
\cite{gmt09}. 

\subsection{Kinetic Equations} 

Given a trajectory of the system, one-point and two-point microscopic
densities in phase space at time $t$ are defined by
\be\label{II.1}
F_1(x_1,t)=\sum_{i=1}^N\delta[x_1-X_i(t)],  
\ee
and
\be\label{II.2}
F_2(x_1,x_2,t)=\sum_{i=1}^N\sum_{j\ne
  i}^N\delta[x_1-X_i(t)]\delta[x_2-X_j(t)], 
\ee
respectively. Here $X_i(t)\equiv\{\mathbf{R}_i(t),\mathbf{V}_i(t)\}$, while
the $x_i\equiv\{\mathbf{r}_i,\mathbf{v}_i\}$ are field variable
referring to the one-particle phase space. The average of $F_1(x_1,t)$ and 
$F_2(x_1,x_2,t)$ over the initial probability distribution of the system
$\rho(\Gamma,0)$, where $\Gamma\equiv\{X_1(t),\dots,X_N(t)\}$, are the usual
one-particle and two-particle distribution functions, 
\begin{equation}\label{distributions}
f_1(x_1,t)=\langle F_1(x_1,t)\rangle, \qquad 
f_2(x_1,x_2,t)=\langle F_2(x_1,x_2,t)\rangle, 
\end{equation}
with the notation 
\begin{equation}
\langle G\rangle\equiv\int d\Gamma G(\Gamma)\rho(\Gamma,0). 
\end{equation}

In the dilute limit, assuming molecular chaos, i.e. that no correlations exist
between colliding particles, and that the sizes of the jumps due to the
thermostat are small compared to the scale on which the distribution varies,
the equation for the one-particle distribution function, $f_1(x_1,t)$, is the
Boltzmann-Fokker-Planck equation \cite{vne98} 
\be\label{I.1}
\frac{\partial}{\partial t}
f_1(x_1,t)+\mathbf{v}_1\cdot\frac{\partial}{\partial\mathbf{r}_1}f_1(x_1,t)
=J[f_1\vert f_1]+\frac{\xi_0^2}{2}\frac{\partial^2}{\partial\mathbf{v}_1^2}
f_1(x_1,t), 
\ee
where 
\be\label{I.2}
J[f_1\vert f_1]=\int dx_2\delta(\mathbf{r}_{12})
\bar{T}_0(\mathbf{v}_1,\mathbf{v}_2)f_1(x_1,t)f_1(x_2,t)
\ee
and
\be\label{I.3}
\bar{T}_0(\mathbf{v}_1,\mathbf{v}_2)=\sigma^{d-1}
\int d\hat{\boldsymbol{\sigma}}
\Theta(\hat{\boldsymbol{\sigma}}\cdot\mathbf{v}_{12})
(\hat{\boldsymbol{\sigma}}\cdot\mathbf{v}_{12})
[\alpha^{-2}b^{-1}_{\hat{\boldsymbol{\sigma}}}-1],
\ee
is the binary collision operator. The operator
$b^{-1}_{\hat{\boldsymbol{\sigma}}}$ changes the 
velocities to its right into the pre-collisional velocities 
\begin{eqnarray}
\mathbf{v}_1^*&=&\mathbf{v}_1-\frac{1+\alpha}{2\alpha}
(\hat{\boldsymbol{\sigma}}\cdot\mathbf{v}_{12})\hat{\boldsymbol{\sigma}},\\
\mathbf{v}_2^*&=&\mathbf{v}_2+\frac{1+\alpha}{2\alpha}
(\hat{\boldsymbol{\sigma}}\cdot\mathbf{v}_{12})\hat{\boldsymbol{\sigma}}.
\end{eqnarray}
As can be seen, the last term in (\ref{I.1}) does not appear in the free
cooling case and depends on the strength of the heating, $\xi_0$.

Let us introduce the two-particle correlation function through the usual
cluster expansion
\begin{equation}\label{correlation}
f_2(x_1,x_2,t)=f_1(x_1,t)f_1(x_2,t)+g_2(x_1,x_2,t). 
\end{equation}
Neglecting three-body correlations, the equation for the correlation function 
$g_2(x_1,x_2,t)$ was derived in \cite{gmt09} and reads
\begin{eqnarray}\label{I.4}
\left[\frac{\partial}{\partial t}
+\mathbf{v}_1\cdot\frac{\partial}{\partial\mathbf{r}_1}
+\mathbf{v}_2\cdot\frac{\partial}{\partial\mathbf{r}_2}\right]g_2(x_1,x_2,t)
&=&\delta(\mathbf{r}_{12})
\bar{T}_0(\mathbf{v}_1,\mathbf{v}_2)f_1(x_1,t)f_1(x_2,t)\nonumber\\
&+&[K(x_1,t)+K(x_2,t)]g_2(x_1,x_2,t)\nonumber\\
&-&\frac{\xi_0^2}{N}
\frac{\partial}{\partial\mathbf{v}_1}\cdot\frac{\partial}{\partial\mathbf{v}_2}
f_1(x_1,t)f_1(x_2,t),\nonumber\\
\end{eqnarray}
where
\be\label{I.5}
K(x_i,t)=\int dx_3\delta(\mathbf{r}_{i3})\bar{T}_0(\mathbf{v}_i,\mathbf{v}_3)
(1+\mathcal{P}_{i3})f_1(x_3,t)+
\frac{\xi_0^2}{2}\frac{\partial^2}{\partial\mathbf{v}_i^2},
\ee
with $\mathcal{P}_{ab}$ an operator that interchanges the label $a$ and $b$ in
the quantities to its right. 

\subsection{The stationary state}

It has been shown numerically that, for a wide class of initial conditions,
the system reaches a homogeneous stationary state \cite{vnetp99} in which the
energy input from the thermostat is compensated by the energy lost in
collisions. In this case the Boltzmann-Fokker-Planck equation reads 
\be\label{I.6}
\frac{\xi_0^2}{2}\frac{\partial^2}{\partial\mathbf{v}_1^2}f_H(\mathbf{v}_1)
+J[f_H\vert f_H]=0.
\ee
For the sake of simplicity, let us introduce the following dimensionless
distribution 
\be\label{I.7}
f_H(\mathbf{v}_1)=\frac{n_H}{v_H^d}\chi_H(c),
\ee
with $\chi_H(c)$ an isotropic function of the modulus of $\mathbf{c}$,
$v_H=(\frac{2T_H}{m})^{1/2}$, $\mathbf{c}=\frac{\mathbf{v}}{v_H}$ and $T_H$
the granular temperature defined as 
\be\label{I.8}
\frac{d}{2}n_HT_H=\int d\mathbf{v}\frac{1}{2}mv^2f_H(\mathbf{v}).
\ee
By introducing equation (\ref{I.7}) in equation (\ref{I.6}), we obtain a
closed equation for  $\chi_H$
\be\label{I.9}
\int d\mathbf{c}_2\bar{T}(\mathbf{c}_1,\mathbf{c}_2)
\chi_H(\mathbf{c}_1)\chi_H(\mathbf{c}_2)
+\frac{\tilde{\xi}^2}{2}\frac{\partial}{\partial\mathbf{c}_1^2}\chi_H(c_1)=0,
\ee
where
\begin{equation}
\bar{T}(\mathbf{c}_1,\mathbf{c}_2)=\int d\hat{\boldsymbol{\sigma}}
\Theta(\hat{\boldsymbol{\sigma}}\cdot\mathbf{c}_{12})(\hat{\boldsymbol{\sigma}}
\cdot\mathbf{c}_{12})[\alpha^{-2}b^{-1}_{\hat{\boldsymbol{\sigma}}}-1],
\end{equation}
is the dimensionless binary collision operator and
$\tilde{\xi}^2=\frac{\xi_0^2\ell}{v_H^3}$ is the dimensionless strength of the
noise with $\ell=(n_H\sigma^{d-1})^{-1}$ proportional to the mean free path. 
In the case of the correlation function, it is convenient to introduce its
dimensionless counterpart, $\tilde{g}_H$, as 
\be\label{I.10}
g_{2,H}(x_1,x_2)=\frac{n_H}{\ell^2v_H^{2d}}
\tilde{g}_H(\mathbf{l}_{12},\mathbf{c}_1,\mathbf{c}_2),
\ee
where we have introduced the dimensionless length scale 
$\mathbf{l}=\mathbf{r}/\ell$ and 
$\mathbf{l}_{12}=\mathbf{l}_1-\mathbf{l}_2$. The reduced distribution fulfills
\begin{eqnarray}\label{I.11}
\left[\Lambda(\mathbf{c}_1)+\Lambda(\mathbf{c}_2)
-\mathbf{c}_{12}\cdot\frac{\partial}{\partial\mathbf{l}_{12}}\right]
\tilde{g}_H(\mathbf{l}_{12},\mathbf{c}_1,\mathbf{c}_2)
&=&-\delta(\mathbf{l}_{12})\bar{T}(\mathbf{c}_1,\mathbf{c}_2)
\chi_H(c_1)\chi_H(c_2)\nonumber\\
&+&\tilde{\xi}^2\frac{n_H\ell^d}{N}
\frac{\partial}{\partial\mathbf{c}_1}\cdot
\frac{\partial}{\partial\mathbf{c}_2}\chi_H(c_1)\chi_H(c_2),\nonumber\\
\end{eqnarray}
where $\Lambda(\mathbf{c}_i)$ is the linearized Boltzmann-Fokker-Planck operator
\be\label{I.12}
\Lambda(\mathbf{c}_i)h(\mathbf{c}_i)=\int d\mathbf{c}_3 
\bar{T}(\mathbf{c}_i,\mathbf{c_3})(1+\mathcal{P}_{i3})
\chi_H(c_3)h(\mathbf{c}_i)+\frac{\tilde{\xi}^2}{2}
\frac{\partial^2}{\partial\mathbf{c}_i^2}h(\mathbf{c}_i).
\ee
Equation (\ref{I.11}) describes the one-time correlation between
fluctuations in the stationary state. As it can be seen, the correlation
function is determined by the properties of the linearized Boltzmann-Fokker-Planck operator,
$\Lambda$, and by the one-particle distribution function, $\chi_H$.

For the purpose of the next section, it is also convenient to define a new
function 
\be\label{II.8}
h_H(\mathbf{l}_{12},\mathbf{c}_1,\mathbf{c}_2)
\equiv \chi_H(c_1)\delta(\mathbf{l}_{12})\delta(\mathbf{c}_{12})
+\tilde{g}_H(\mathbf{l}_{12},\mathbf{c}_1,\mathbf{c}_2).
\ee
Taking into account equation (\ref{I.11}) and that the term
$\mathbf{c}_{12}\cdot\frac{\partial}{\partial\mathbf{l}_{12}}\chi_H(c_1)
\delta(\mathbf{l}_{12})\delta(\mathbf{c}_{12})$
vanishes identically, the equation for this quantity is  
\begin{eqnarray}\label{II.9}
&&\left[\Lambda(\mathbf{c}_1)+\Lambda(\mathbf{c}_2)
-\mathbf{c}_{12}\cdot\frac{\partial}{\partial \mathbf{l}_{12}}\right]
h_H(\mathbf{l}_{12},\mathbf{c}_1,\mathbf{c}_2)\nonumber\\
&&\qquad\qquad=-\delta(\mathbf{l}_{12})\Gamma(\mathbf{c}_1,\mathbf{c}_2)
+\tilde{\xi}^2\frac{n_H\ell^d}{N}
\frac{\partial}{\partial\mathbf{c}_1}\cdot\frac{\partial}{\partial\mathbf{c}_2}
\chi_H(c_1)\chi_H(c_2),
\end{eqnarray}
where
\be\label{II.10}
\Gamma(\mathbf{c}_1,\mathbf{c}_2)=\bar{T}(\mathbf{c}_1,\mathbf{c}_2)
\chi_H(c_1)\chi_H(c_2)-[\Lambda(\mathbf{c}_1)+\Lambda(\mathbf{c}_2)]
\chi_H(c_1)\delta(\mathbf{c}_{12}).
\ee

\subsection{Linearized Boltzmann-Fokker-Planck equation}

In this section we derive the evolution equation for a small perturbation
around the homogeneous stationary distribution function $f_H(\mathbf{v})$. The
objective is to evaluate some spectral properties of the linear operator that
controls the dynamics, which turns out to be the linearized Boltzmann-Fokker-Planck
operator introduced above. These properties 
will be useful in the last section where we derive hydrodynamic equations. 

Let us introduce the small perturbation, $\delta f$ 
\be\label{I.13}
\delta f(x_1,t)=f_1(x_1,t)-f_H(\mathbf{v}_1),\qquad\frac{\delta f}{f_1}\ll 1.
\ee
The linearized equation for $\delta f$ is obtained from the Boltzmann-Fokker-Planck equation
(\ref{I.1}) and equation (\ref{I.6})
\be\label{I.14}
\frac{\partial}{\partial t}\delta f(x_1,t)
+\mathbf{v}_1\cdot\frac{\partial}{\partial\mathbf{r}_1} \delta f(x_1,t)
=K(x_1,t)\delta f(x_1,t),
\ee 
where $K(x_1,t)$ is defined in (\ref{I.5}). Now, let us introduce a
dimensionless perturbation and a dimensionless time scale as 
\be\label{I.15}
\delta f(x_1,t)=\frac{n_H}{v_H^d}\delta\chi(\mathbf{l},\mathbf{c},s),
\ee
\be\label{I.16}
s=\frac{v_H}{\ell}t,
\ee
which is proportional to the number of collisions per particle in the interval
$(0,t)$. The equation for $\delta\chi$ reads
\be\label{I.17}
\frac{\partial}{\partial s}\delta\chi(\mathbf{l},\mathbf{c}_1,s)
=\left[\Lambda(\mathbf{c}_1)
-\mathbf{c}_1\cdot\frac{\partial}{\partial\mathbf{l}}\right]
\delta\chi(\mathbf{l},\mathbf{c}_1,s),
\ee
where $\Lambda$ is the linearized Boltzmann-Fokker-Planck operator defined in
(\ref{I.12}). Equation (\ref{I.17}) is the linearized Boltzmann-Fokker-Planck equation and
it describes the dynamics of any small perturbation around the homogeneous
stationary state. 

In \cite{gmt09} it was shown that the linearized Boltzmann-Fokker-Planck operator has $d+1$
eigenfunctions associated to the null eigenvalue (in principle there are not
more eigenfunctions associated to this eigenvalue). These eigenfunctions are
\be\label{I.18}
\xi_1(\mathbf{c})
=\frac{1}{3}\frac{\partial}{\partial\mathbf{c}}\cdot[\mathbf{c}\chi_H(c)]
+\chi_H(c),
\ee
\be\label{I.19}
\boldsymbol{\xi}_2(\mathbf{c})=-\frac{\partial}{\partial\mathbf{c}}\chi_H(c).
\ee
They fulfill $\Lambda(\mathbf{c})\xi_i(\mathbf{c})=0$ for $i=1,2$. Moreover,
as the number of particles and momentum are conserved in collisions, we have
\be\label{I.20}
\int d\mathbf{c}\Lambda(\mathbf{c})h(\mathbf{c})
=\int d\mathbf{c} c_i\Lambda(\mathbf{c})h(\mathbf{c})=0,
\ee
where $h$ is an arbitrary function.
Equivalently, it can be said that the functions
\be\label{I.21}
\bar{\xi}_1(\mathbf{c})=\chi_H(c),
\qquad\bar{\boldsymbol{\xi}}_2(\mathbf{c})=\chi_H(c)\mathbf{c},
\ee
are left eigenfunctions of $\Lambda$ associated to the null eigenvalue
with the scalar product 
\be\label{I.22}
\langle f\vert g\rangle=
\int d\mathbf{c}\chi_H^{-1}(c)f^*(\mathbf{c})g(\mathbf{c}),
\ee
with $f^*$ the complex conjugate of $f$.

\section{The Boltzmann-Langevin equation}\label{ecboltzmannlangevin}

The starting point for the study of the fluctuations in this work will be the
Boltzmann-Langevin equation. In this section we derive the corresponding
equation and we determine the properties of the noise in order to obtain
consistency with the equations of the correlation functions presented in the
previous section.

As defined in Eq. (\ref{distributions}), the one-particle distribution
function, $f_1$ is the ensemble average of the phase function $F_1$, and its
dynamics is given by the Boltzmann-Fokker-Planck equation, Eq. (\ref{I.1}). 
The problem
is now to find an evolution equation for the fluctuating quantity
\be\label{II.3}
\delta
F(\mathbf{l},\mathbf{c},s)=\frac{v_H^d}{n_H}[F_1(x,t)-f_H(\mathbf{v})]. 
\ee
As for the velocity of a Brownian particle \cite{resibois}, we expect 
that the difference between the  equation for the fluctuating quantity $\delta
F$,  and its average, $\langle\delta F\rangle=\delta\chi$, is a random force
term, $R$ \cite{bz69}. Then, the fluctuations $\delta F$ around $\chi_H$ are
described by a Boltzmann equation linearized around the solution $\chi_H$ with 
a random force, $R$, added
\be\label{II.4}
\frac{\partial}{\partial s}\delta F(\mathbf{l},\mathbf{c},s)
=\left[\Lambda(\mathbf{c})
-\mathbf{c}\cdot\frac{\partial}{\partial\mathbf{l}}\right]
\delta F(\mathbf{l},\mathbf{c},s)+R(\mathbf{l},\mathbf{c},s). 
\ee
Taking averages
in equation (\ref{II.4}), we obtain
the linearized Boltzmann-Fokker-Planck equation, Eq. (\ref{I.17}), if and only if 
\be
\langle R(\mathbf{l},\mathbf{c},s)\rangle=0. 
\ee
Equation (\ref{II.4}) is the Boltzmann-Langevin equation. As in the 
free cooling case \cite{bmg09}, we assume that the noise term, 
$R(\mathbf{l},\mathbf{c},s)$, is Markovian
\be\label{II.5}
\langle R(\mathbf{l}_1,\mathbf{c}_1,s_1)
R(\mathbf{l}_2,\mathbf{c}_2,s_2)\rangle_H
=H(\mathbf{l}_1,\mathbf{l}_2,\mathbf{c}_1,\mathbf{c}_2)\delta(s_1-s_2),
\ee
where $\langle\dots\rangle_H$ means average in the stationary homogeneous
state. 
In order to evaluate the function 
$H(\mathbf{l}_1,\mathbf{l}_2,\mathbf{c}_1,\mathbf{c}_2)$ explicitely, we will
calculate $\langle\delta F(\mathbf{l}_1,\mathbf{c}_1,s)
\delta F(\mathbf{l}_2,\mathbf{c}_2,s)\rangle_H$
with the Boltzmann-Langevin equation and then, we will impose compatibility
with the equation of the correlation function of the previous section. 
So, let us first write this function as a functional of the
distribution and the correlation functions. Using the definitions of the
microscopic densities, Eqs. (\ref{II.1}), (\ref{II.2}), the distribution and
correlation functions, Eqs. (\ref{distributions}), (\ref{correlation}),  
(\ref{II.8}), and the dimensionless distributions, Eqs. (\ref{I.7}), 
(\ref{I.10}), we have  
\begin{eqnarray}\label{II.7}
\langle\delta F(x_1,t)\delta F(x_2,t)\rangle_H
&=&\frac{v_H^{2d}}{n_H^2}\langle[F_1(x_1,t)-f_H(\mathbf{v}_1)]
[F_1(x_2,t)-f_H(\mathbf{v}_2)]\rangle_H\nonumber\\
&=&\frac{v_H^ {2d}}{n_H^2}
[\langle F_1(x_1,t)F_1(x_2,t)\rangle_H-f_H(\mathbf{v}_1)f_H(\mathbf{v}_2)]
\nonumber\\
&=&\frac{v_H^{2d}}{n_H^2}[f_{2,H}(x_1,x_2,t)
+f_H(\mathbf{v}_1)\delta(x_1-x_2)-f_H(\mathbf{v}_1)f_H(\mathbf{v}_2)]
\nonumber\\
&=&\frac{v_H^{2d}}{n_H^2}[g_{2,H}(x_1,x_2)+f_H(\mathbf{v}_1)\delta(x_1-x_2)]
\nonumber\\
&=&\frac{1}{n_H\ell^d}[\tilde{g}_H(\mathbf{l}_{12},\mathbf{c}_1,\mathbf{c}_2)
+\delta(\mathbf{l}_{12})\delta(\mathbf{c}_{12})\chi_H(c_1)]
\nonumber\\
&=&\frac{1}{n_H\ell^d}h_H(\mathbf{l}_{12},\mathbf{c}_1,\mathbf{c}_2).
\end{eqnarray}

Now, let us solve Eq. (\ref{II.4}) as a functional of the noise. In order to
do that it is convenient to define the linear operator
\be
\Lambda(\mathbf{l}_i,\mathbf{c}_i)\equiv\Lambda(\mathbf{c}_i)
-\mathbf{c}_i\cdot\frac{\partial}{\partial\mathbf{l}_i}, 
\ee
in terms of which the solution for $\delta F(\mathbf{l},\mathbf{c},s)$ is
\begin{eqnarray}\label{II.11}
\delta F(\mathbf{l},\mathbf{c},s)
&=&e^{\Lambda(\mathbf{l},\mathbf{c})s}
\delta F(\mathbf{l},\mathbf{c},0)+\int_0^s ds^\prime 
e^{\Lambda(\mathbf{l},\mathbf{c})(s-s')}
R(\mathbf{l},\mathbf{c},s^\prime)\nonumber\\
&&\stackrel{s\gg 1}{\to}
\int_0^s ds^\prime e^{\Lambda(\mathbf{l},\mathbf{c})(s-s^\prime)}
R(\mathbf{l},\mathbf{c},s^\prime), 
\end{eqnarray}
where we have assumed that the term stemming from 
the initial conditions vanishes in
the long time limit. This is equivalent to assuming that the spectrum of
the linearized Boltzmann-Fokker-Planck operator is such that any perturbation without
component in the subspace associated to the null eigenvalue decays. With
equation (\ref{II.11}), we can evaluate 
\begin{eqnarray}\label{II.12}
&&\langle \delta F(\mathbf{l}_1,\mathbf{c}_1,s)
\delta F(\mathbf{l}_2,\mathbf{c}_2,s)\rangle_H\nonumber\\
&&=\int_0^s ds^\prime\int_0^s ds^{\prime\prime}
e^{\Lambda(\mathbf{l}_1,\mathbf{c}_1)(s-s^\prime)
+\Lambda(\mathbf{l}_2,\mathbf{c}_2)(s-s^{\prime\prime})}
\langle R(\mathbf{l}_1,\mathbf{c}_1,s^\prime)
R(\mathbf{l}_2,\mathbf{c}_2,s^{\prime\prime})\rangle_H\nonumber\\
&&=\int_0^s ds^\prime\int_0^s ds^{\prime\prime}
e^{\Lambda(\mathbf{l}_1,\mathbf{c}_1)(s-s^\prime)
+\Lambda(\mathbf{l}_2,\mathbf{c}_2)(s-s^{\prime\prime})}
H(\mathbf{l}_{12},\mathbf{c}_1,\mathbf{c}_2)\delta(s^{\prime}-s^{\prime\prime})
\nonumber\\
&&=\int_0^s ds^\prime e^{[\Lambda(\mathbf{l}_1,\mathbf{c}_1)
+\Lambda(\mathbf{l}_2,\mathbf{c}_2)](s-s^{\prime})}
H(\mathbf{l}_{12},\mathbf{c}_1,\mathbf{c}_2)\nonumber\\
&&=-[\Lambda(\mathbf{l}_1,\mathbf{c}_1)
+\Lambda(\mathbf{l}_2,\mathbf{c}_2)]^{-1}
\left[e^{[\Lambda(\mathbf{l}_1,\mathbf{c}_1)
+\Lambda(\mathbf{l}_2,\mathbf{c}_2)]
(s-s^\prime)}\right]_{s^\prime=0}^{s^\prime=s}
H(\mathbf{l}_{12},\mathbf{c}_1,\mathbf{c}_2)\nonumber\\
&&\stackrel{s\gg 1}{\to}-[\Lambda(\mathbf{l}_1,\mathbf{c}_1)
+\Lambda(\mathbf{l}_2,\mathbf{c}_2)]^{-1}
H(\mathbf{l}_{12},\mathbf{c}_1,\mathbf{c}_2),
\end{eqnarray}
where we have assumed that the term 
$e^{[\Lambda(\mathbf{l}_1,\mathbf{c}_1)+\Lambda(\mathbf{l}_2,\mathbf{c}_2)]s}
H(\mathbf{l}_{12},\mathbf{c}_1,\mathbf{c}_2)\to 0$ in the long time limit. 
After identifying the function, we will see that this is, in fact, the case,
because $H(\mathbf{l}_{12},\mathbf{c}_1,\mathbf{c}_2)$ does not have 
components in the subspace associated to the null eigenvalue. 
Equivalently we have
\be\label{II.13}
[\Lambda(\mathbf{l}_1,\mathbf{c}_1)+\Lambda(\mathbf{l}_2,\mathbf{c}_2)]
h_H(\mathbf{l}_{12},\mathbf{c}_1,\mathbf{c}_2)
=-n_H\ell^dH(\mathbf {l}_{12},\mathbf{c}_1,\mathbf{c}_2).
\ee 
Finally, comparing equations (\ref{II.13}) and (\ref{II.9}) we conclude that
\be\label{II.14}
H(\mathbf{l}_{12},\mathbf{c}_1,\mathbf{c}_2)
=\frac{1}{n_H\ell^d}\delta(\mathbf{l}_{12})
\Gamma(\mathbf{c}_1,\mathbf{c}_2)-\frac{\tilde{\xi}^2}{N}
\frac{\partial}{\partial\mathbf{c}_1}\cdot
\frac{\partial}{\partial\mathbf{c}_2}\chi_H(c_1)\chi_H(c_2), 
\ee
with $\Gamma(\mathbf{c}_1,\mathbf{c}_2)$ given in Eq. (\ref{II.10}).  With
this expression of $H$ we can see that it does not have components in the 
subspace associated to the null eigenvalue. Taking into account that 
$\bar{\xi}_1$ and $\bar{\boldsymbol{\xi}}_2$ are left eigenfunctions of 
$\Lambda$  associated to the null eigenvalue and that \cite{gmt09}
\begin{eqnarray}
\int d\mathbf{c}_1\int d\mathbf{c}_2\bar{T}(\mathbf{c}_1,\mathbf{c}_2)
\chi_H(\mathbf{c}_1)\chi_H(\mathbf{c}_2)=0, \nonumber\\
\int d\mathbf{c}_1\int d\mathbf{c}_2c_{1i}c_{2i}
\bar{T}(\mathbf{c}_1,\mathbf{c}_2)\chi_H(\mathbf{c}_1)\chi_H(\mathbf{c}_2)=
\tilde{\xi}^2, 
\end{eqnarray}
we can see that
\begin{eqnarray}
\int d\mathbf{l}_{12}\int d\mathbf{c}_1\int d\mathbf{c}_2
H(\mathbf {l}_{12},\mathbf{c}_1,\mathbf{c}_2)=0,\nonumber\\
\int d\mathbf{l}_{12}\int d\mathbf{c}_1\int d\mathbf{c}_2c_{1i}c_{2i}
H(\mathbf {l}_{12},\mathbf{c}_1,\mathbf{c}_2)=0. 
\end{eqnarray}

In the remaining of this section, we shall write the Boltzmann-Langevin
equation together with noise properties in Fourier space. This will prove
useful for the subsequent analysis. Let us introduce the Fourier component of a
function of the position variable as
\be\label{II.15}
\tilde{f}(\mathbf{k})=\int d\mathbf{l}e^{-i\mathbf{k}\cdot\mathbf{l}}
f(\mathbf{l}),
\qquad f(\mathbf{l})=\frac{1}{\tilde{V}}\sum_{\mathbf{k}}
e^{i\mathbf{k}\cdot\mathbf{l}}\tilde{f}(\mathbf{k}),
\ee 
where $\tilde{V}=\frac{V}{\ell^d}$ is the volume in units of the mean free
path. The equation for $\delta\tilde{F}(\mathbf{k},\mathbf{c},s)$ is then 
\be\label{II.16}
\left[\frac{\partial}{\partial s}-\Lambda(\mathbf{k},\mathbf{c})\right]
\delta\tilde{F}(\mathbf{k},\mathbf{c},s)=\tilde{R}(\mathbf{k},\mathbf{c},s), 
\ee
where we have introduced the operator
\be
\Lambda(\mathbf{k},\mathbf{c})=\Lambda(\mathbf{c})-i\mathbf{k}\cdot\mathbf{c}. 
\ee
The Fourier transform of the noise, $\tilde{R}(\mathbf{k},\mathbf{c},s)$, obeys
\be
\langle \tilde{R}(\mathbf{k},\mathbf{c},s)\rangle_H=0, 
\ee
and 
\be\label{II.17}
\langle \tilde{R}(\mathbf{k}_1,\mathbf{c}_1,s_1)
\tilde{R}(\mathbf{k}_2,\mathbf{c}_2,s_2)\rangle_H=
H(\mathbf{k}_1,\mathbf{k}_2,\mathbf{c}_1,\mathbf{c}_2)\delta(s_1-s_2),
\ee
where
\be\label{II.18}
H(\mathbf{k}_1,\mathbf{k}_2,\mathbf{c}_1,\mathbf{c}_2)
=\frac{\tilde{V}^2}{N}\left[\Gamma(\mathbf{c}_1,\mathbf{c}_2)
\delta(\mathbf{k}_1+\mathbf{k}_2)-\tilde{\xi}^2
\frac{\partial}{\partial \mathbf{c}_1}\cdot
\frac{\partial}{\partial\mathbf{c}_2}
\chi_H(c_1)\chi_H(c_2)\delta(\mathbf{k}_1)\delta(\mathbf{k}_2)\right],
\ee
which completes the calculation of the noise variance within the 
Langevin description, that will be used to quantify the 
transverse velocity fluctuations. 

\section{The fluctuating transverse velocity}
\label{transversevelocity}

The objective in this section is to derive a fluctuating equation for the
transverse velocity field, $\mathbf{w}_\perp(\mathbf{k},s)$. The reason to
consider this field is that its equation is decoupled from the equations
for the other hydrodynamic fields \cite{vnetp99}, and it can be derived
exactly in the hydrodynamic limit with the knowledge we have about the
spectrum of the linearized Boltzmann-Fokker-Planck operator.  

Mathematically, the transverse velocity is defined in the following way: Let
us consider the $d$-dimensional vector $\mathbf{k}$ which belongs to the space
$\Re^d$. This space can be expanded in the base
$\left\{\mathbf{\hat{k}}\right\} 
\bigcup\left\{\mathbf{\hat{k}}^i_\perp\right\}_{i=1}^{d-1}$
where $\hat{\mathbf{k}}$ is a unitary vector parallel to $\mathbf{k}$ and
$\left\{\mathbf{\hat{k}}^i_\perp\right\}_{i=1}^{d-1}$ are $(d-1)$ unitary
vectors orthogonal to $\mathbf{k}$. The transverse velocity is defined as
\be\label{III.1}
w_{\perp i}(\mathbf{k},s)
=\int d\mathbf{c}(\mathbf{c}\cdot\mathbf{\hat{k}}_\perp^i)
\delta\tilde{F}(\mathbf{k},\mathbf{c},s), \qquad i=1,\dots,d-1.
\ee
For the subsequent analysis, is is convenient to introduce the following
projectors  
\be\label{III.2a}
P^{(i)}f(\mathbf{c})\equiv\langle\bar{\xi}_{2\perp i}(\mathbf{c})\vert 
f(\mathbf{c})\rangle\xi_{2\perp i}(\mathbf{c}),
\ee
\be\label{III.2b}
Q^{(i)}f(\mathbf{c})\equiv(1-P^{(i)})f(\mathbf{c}).
\ee
Then, if we apply $P^{(i)}$ to $\delta \tilde{F}$ we obtain
\be\label{III.3}
P^{(i)}\delta\tilde{F}(\mathbf{k},\mathbf{c},s)
=w_{\perp i}(\mathbf{k},s)\xi_{2\perp i}(\mathbf{c}),
\ee
and the transverse velocity is the component of 
$\delta\tilde{F}(\mathbf{k},\mathbf{c},s)$ in the subspace generated by 
$\xi_{2\perp i}(\mathbf{c})$. 

In order to obtain an equation for $\mathbf{w}_\perp$, we apply the projectors
$P$ and $Q$ (for simplicity in the notation we skip the super-index) to the
Langevin equation (\ref{II.16})  
\be\label{III.4}
\left[\frac{\partial}{\partial s}-P\Lambda(\mathbf{k},\mathbf{c})\right]
P\delta\tilde{F}(\mathbf{k},\mathbf{c},s)=P\tilde{R}(\mathbf{k},\mathbf{c},s)
-Pi(\mathbf{k}\cdot\mathbf{c})Q\delta\tilde{F}(\mathbf{k},\mathbf{c},s),
\ee
\be\label{III.5}
\left[\frac{\partial}{\partial s}-Q\Lambda(\mathbf{k},\mathbf{c})\right]
Q\delta\tilde{F}(\mathbf{k},\mathbf{c},s)=Q\tilde{R}(\mathbf{k},\mathbf{c},s)
-Qi(\mathbf{k}\cdot\mathbf{c})P\delta\tilde{F}(\mathbf{k},\mathbf{c},s).
\ee
Now, let us solve equation (\ref{III.5}) formally   
\begin{eqnarray}\label{III.6}
Q\delta\tilde{F}(\mathbf{k},\mathbf{c},s)
&=&e^{Q\Lambda(\mathbf{k},\mathbf{c})s}Q
\delta\tilde{F}(\mathbf{k},\mathbf{c},0)\nonumber\\
&+&\int_0^s ds^\prime e^{Q\Lambda(\mathbf{k},\mathbf{c})(s-s^\prime)}
[Q\tilde{R}(\mathbf{k},\mathbf{c},s^\prime)-Qi(\mathbf{k}\cdot\mathbf{c})
P\delta\tilde{F}(\mathbf{k},\mathbf{c},s^\prime)].\qquad
\end{eqnarray}
In the time regime in which the system has forgotten the initial
  condition, i.e. when we can consider that $e^{Q\Lambda(\mathbf{k},\mathbf{c})s}Q
\delta\tilde{F}(\mathbf{k},\mathbf{c},0)\to 0$, 
by substituting Eq. (\ref{III.6}) in Eq. (\ref{III.4}), we
obtain a closed equation for $P\delta\tilde{F}(\mathbf{k},\mathbf{c},s)$ 
\begin{eqnarray}\label{III.7}
\left[\frac{\partial}{\partial s}-P\Lambda(\mathbf{k},\mathbf{c})\right]
P\delta\tilde{F}(\mathbf{k},\mathbf{c},s)
&+&P(\mathbf{k}\cdot\mathbf{c})\int_0^s ds^\prime 
e^{Q\Lambda(\mathbf{k},\mathbf{c})(s-s^\prime)}
Q(\mathbf{k}\cdot\mathbf{c})P\delta\tilde{F}(\mathbf{k},\mathbf{c},s^\prime)
\nonumber\\
&=&P\tilde{R}(\mathbf{k},\mathbf{c},s)\nonumber\\
&-&Pi(\mathbf{k}\cdot\mathbf{c})\int_0^s ds^\prime 
e^{Q\Lambda(\mathbf{k},\mathbf{c})(s-s^\prime)}
Q\tilde{R}(\mathbf{k},\mathbf{c},s^\prime).
\end{eqnarray}
In Appendix A it is shown that, in the hydrodynamic limit, i.e. to second order
in $k$ and in the long time limit, equation (\ref{III.7}) reduces to the
following equation for the transverse velocity
\be\label{III.8}
\left[\frac{\partial}{\partial
    s}+\tilde{\eta}k^2\right]w_{\perp}(\mathbf{k},s)
=R_w(\mathbf{k},s). 
\ee
The coefficient $\tilde{\eta}$ is the shear viscosity given by
\be\label{III.9}
\tilde{\eta}=\int^\infty_0 ds\int d\mathbf{c} c_x c_y 
e^{\Lambda(\mathbf{c}) s}c_x\xi_{2,y}(\mathbf{c})=-\int d\mathbf{c} c_x c_y 
\Lambda(\mathbf{c})^{-1}c_x\xi_{2,y}(\mathbf{c}),
\ee
which agrees with the expression obtained in \cite{gm02} by the Chapman-Enskog
method, and $R_w(\mathbf{k},s)$ is a noise term which can be decomposed as
\be
R_w(\mathbf{k},s)=S(\mathbf{k},s)+\Pi(\mathbf{k},s). 
\ee
The term $S(\mathbf{k},s)$ comes from the thermostat (which does
not conserve momentum locally), and the
second term, $\Pi(\mathbf{k},s)$, is the fluctuating part of the
pressure tensor. Their 
microscopic expressions in terms of the noise of the Boltzmann-Langevin
equation are
\be\label{III.10}
S(\mathbf{k},s)=\int d\mathbf{c}(\mathbf{\hat{k}_\perp}\cdot\mathbf{c})
\tilde{R}(\mathbf{k},\mathbf{c},s),
\ee
and
\be\label{III.11}
\Pi(\mathbf{k},s)=-ik\int_0^s ds^\prime\int 
d\mathbf{c}(\mathbf{\hat{k}}\cdot\mathbf{c})(\mathbf{\hat{k}_\perp}
\cdot\mathbf{c})e^{Q\Lambda(\mathbf{k},\mathbf{c})(s-s^\prime)}
Q\tilde{R}(\mathbf{k},\mathbf{c},s^\prime).
\ee
Of course, due to the fact that 
$\langle\tilde{R}(\mathbf{k},\mathbf{c},s)\rangle_H=0$, the mean value of the
noise vanishes
\be\label{mean_value}
\langle R_w(\mathbf{k},s)\rangle_H=0.
\ee
The autocorrelation function of the noise is evaluated in Appendix B. Due to
symmetry considerations, there are only correlations between the $\mathbf{k}$
and $-\mathbf{k}$ components, which yields
\begin{eqnarray}\label{III.12}
\langle R_w(\mathbf{k},s_1)R_w(-\mathbf{k},s_2)\rangle_H
&=&\frac{\tilde{V}^2}{N}\left[\tilde{\xi}^2\delta(s_1-s_2)
+k^2C_{xy}(s_2-s_1)\right]\nonumber\\
&+&\langle S(\mathbf{k},s_1)\Pi(-\mathbf{k},s_2)\rangle_H, \qquad s_1< s_2. 
\end{eqnarray}
The first term is the expected zeroth order term which comes from the
heating. The Dirac delta function is an exact consequence of the fact that the
external noise (the heating) is white. 
The function $C_{xy}(s_2-s_1)$ reads
\begin{equation}
C_{xy}(s_2-s_1)=\int d\mathbf{c}_1\int d\mathbf{c}_2 
c_{1x}c_{1y}c_{2x}c_{2y}
e^{\Lambda(\mathbf{c}_2)(s_2-s_1)}\phi_H(\mathbf{c}_1,\mathbf{c}_2). 
\end{equation}
Here, we have introduced the function $\phi_H(\mathbf{c}_1,\mathbf{c}_2)$ as
the space integral of the 
correlation function $h_H(\mathbf{l},\mathbf{c}_1,\mathbf{c}_2)$
\be\label{def_phi}
\phi_H(\mathbf{c}_1,\mathbf{c}_2)
=\int d\mathbf{l}h_H(\mathbf{l},\mathbf{c}_1,\mathbf{c}_2)=
\chi_H(c_1)\delta(\mathbf{c}_{12})+\chi_2(\mathbf{c}_1,\mathbf{c}_2), 
\ee
where $\chi_2(\mathbf{c}_1,\mathbf{c}_2)
=\int d\mathbf{l}\tilde{g}_H(\mathbf{l},\mathbf{c}_1,\mathbf{c}_2)$ and 
$\tilde{g}_H$ is the dimensionless correlation function defined in
(\ref{I.10}). The equation for $\phi_H$ can easily be 
obtained by integration over space variable in Eq. (\ref{II.9}),
which gives 
\begin{equation}\label{ec_phi}
[\Lambda(\mathbf{c}_1)+\Lambda(\mathbf{c}_2)]\phi_H(\mathbf{c}_1,\mathbf{c}_2)
=-\Gamma(\mathbf{c}_1,\mathbf{c}_2)+\tilde{\xi}^2
\frac{\partial}{\partial \mathbf{c}_1}\cdot
\frac{\partial}{\partial\mathbf{c}_2}\chi_H(c_1)\chi_H(c_2).
\end{equation}
As discussed in Appendix B, $C_{xy}$ can be physically interpreted as the
autocorrelation function of the global quantity  
$\sum_{i=1}^NV_x(t)V_y(t)$. In the elastic case, this correlation function is
related to the shear viscosity but it is not the case for granular systems
\cite{bdb08}. The formula for the correlation 
$\langle S(\mathbf{k},s_1)\Pi(-\mathbf{k},s_2)\rangle_H$ is given in Appendix
B, and it does not seem to admit a direct physical interpretation. 
As the two correlation functions, $C_{xy}(s)$ and 
$\langle S(\mathbf{k},s_1)\Pi(-\mathbf{k},s_2)\rangle_H$, decay with the
kinetic modes and, in the $k\to 0$ limit, the fluctuating velocity is
expected to be frozen (its time evolution is given by the null
  eigenvalue, Eq. (\ref{III.8})), we can consider that they are proportional to a Dirac
delta function in time, and we have
\begin{eqnarray}\label{III.13}
&&\frac{\tilde{V}^2}{N}k^2C_{xy}(s_2-s_1)
+\langle S(\mathbf{k},s_1)\Pi(-\mathbf{k},s_2)\rangle_H\nonumber\\
&&\to 2\left[\frac{\tilde{V}^2}{N}k^2\int_0^\infty dsC_{xy}(s)
+\int_0^\infty ds \langle S(\mathbf{k},0)\Pi(-\mathbf{k},s)\rangle_H\right]
\delta(s_1-s_2).\nonumber\\
\end{eqnarray}
Note that this is in contrast with the free cooling case where the noise 
can not be considered to be white \cite{bmg09}. In this case, the equation for
the transverse velocity (rescaled with the thermal velocity) contains a term
of order zero in $k$ which is proportional to the cooling rate. Then, even in
the $k\to 0$ limit, the velocity is not frozen on the kinetic scale. In
Appendix C the second integral of equation (\ref{III.13}) is 
evaluated obtaining 
\be\label{III.14}
\int_0^\infty ds \langle S(\mathbf{k},0)\Pi(-\mathbf{k},s)\rangle_H
\to\frac{\tilde{V}^2}{N}k^2\int d\mathbf{c}_1\int d\mathbf{c}_2 
c_{1x}c_{2x}c_{2y}\Lambda(\mathbf{c}_2)^{-1}c_{2y}
\phi_H(\mathbf{c}_1,\mathbf{c}_2), 
\ee
which is valid in the hydrodynamic limit. 
If we substitute $\phi_H(\mathbf{c}_1,\mathbf{c}_2)$ by its expression in
terms of the one and two-particle distribution function, Eq. (\ref{def_phi}),
we obtain (see Appendix D) that the one-particle contribution vanishes and the 
correlation function can be written in terms of the two-particle velocity
correlation function, $\chi_2(\mathbf{c}_1,\mathbf{c}_2)$
\begin{eqnarray}\label{III.16}
&&\langle R_w(\mathbf{k},s_1)R_w(-\mathbf{k},s_2)\rangle_H\nonumber\\
&&=\frac{\tilde{V}^2}{N}\delta(s_1-s_2)\left\{\tilde{\xi}^2
+2 k^2\left[-\int d\mathbf{c}_1\int d\mathbf{c}_2 
c_{1x}c_{1y}c_{2x}c_{2y}
\Lambda(\mathbf{c}_2)^{-1}\chi_2(\mathbf{c}_1,\mathbf{c}_2)\right.
\right.\nonumber\\
&&\qquad\quad\qquad\qquad\qquad\left.\left.
+\int d\mathbf{c}_1\int d\mathbf{c}_2 c_{1x}c_{2x}c_{2y}ç
\Lambda(\mathbf{c}_2)^{-1}c_{2y}
\chi_2(\mathbf{c}_1,\mathbf{c}_2)\right]\right\}.
\end{eqnarray}
As $R_w(\mathbf{k},s_1)$ is Gaussian, the noise is completely
characterized by Equations (\ref{mean_value}) and (\ref{III.16}). As it can
be seen, the $k^2$ term has no relation, a priori, with the shear viscosity,
Eq. (\ref{III.9}), i.e. there is a priori no 
fluctuation-dissipation relation as that
assumed in \cite{vnetp99}. {\em However}, we now show that, under additional
hypothesis (that in principle are not restricted to the elastic limit), the
aforementioned term reduces to the shear viscosity. 
Let us assume that the most important contribution of the two particle
velocity correlation function, $\chi_2(\mathbf{c}_1,\mathbf{c}_2)$, is the
hydrodynamic part, i.e. we assume
\be\label{approx}
\chi_2(\mathbf{c}_1,\mathbf{c}_2)\simeq
\sum_{\beta=1}^{d+2}\sum_{\beta'=1}^{d+2}a_{\beta,\beta'}
\xi_\beta(\mathbf{c}_1)\xi_{\beta'}(\mathbf{c}_2). 
\ee
This assumption was already made in \cite{gmt09} where the coefficients
$a_{\beta,\beta'}$ were evaluated to calculate the total energy fluctuations.
We emphasize that it led to an excellent agreement between analytical
predictions and numerical data (Monte Carlo) for energy fluctuations,
{\rm for all the values of the inelasticity} \cite{gmt09}. In this
approximation, the first integral in (\ref{III.16}) vanishes because
$c_xc_y\chi_H(c)$ is orthogonal to the hydrodynamic modes. The second term is 
\begin{eqnarray}
\int d\mathbf{c}_1\int d\mathbf{c}_2 c_{1x}c_{2x}c_{2y}
\Lambda(\mathbf{c}_2)^{-1}c_{2y}
\sum_{\beta=1}^{d+2}\sum_{\beta'=1}^{d+2}a_{\beta,\beta'}
\xi_\beta(\mathbf{c}_1)\xi_{\beta'}(\mathbf{c}_2)&&\nonumber\\
=a_{2x,2x}\int d\mathbf{c}_1c_{1x}\xi_{2x}(\mathbf{c}_1)
\int d\mathbf{c}_2 c_{2x}c_{2y}\Lambda(\mathbf{c}_2)^{-1}
c_{2y}\xi_{2x}(\mathbf{c}_2), 
\end{eqnarray}
where, for symmetry considerations, the only
term that remains is the one associated to $\beta=\beta'=2$. If we use now that
$a_{2i2i}=-1/2$ (see the reference \cite{gmt09}), and the formula for the
shear viscosity, Eq. (\ref{III.9}), we finally have  
\be\label{III.17}
\langle R_w(\mathbf{k},s_1) R_w(-\mathbf{k},s_2)
=\frac{\tilde{V}^2}{N}\delta(s_1-s_2)(\tilde{\xi}^2+\tilde{\eta}k^2).
\ee
Then, in the hydrodynamic limit and assuming that the two-particle velocity
correlation function, $\chi_2(\mathbf{c}_1,\mathbf{c}_2)$, has only
components in the hydrodynamic subspace, the correlation function of the noise
reduces to the one introduced phenomenologically in \cite{vnetp99}, not only
in the elastic limit, but for arbitrary inelasticity. Although we have no
direct proof of the accuracy of approximation (\ref{approx}), we note that 
it is backed up by numerical data, see e.g. \cite{gmt09}.

\section{Conclusions and outlook}\label{conclusions}

The primary objective of this work was to derive a Boltzmann-Langevin
description for a heated granular gas, in the spirit of the work of Bixon and
Zwanzig \cite{bz69}, who considered conservative
systems. The system considered here is on the other hand dissipative,
and is heated by a stochastic force changing the
velocity of the particles between collisions. The loss of energy in collisions
is compensated by the energy given to the particles  by the thermostat and a
stationary state is thereby reached. Our study is restricted to this stationary
state. The Boltzmann- Langevin equation has been derived and the properties of
the noise appearing in this equation were identified by imposing consistency
with the equation for the correlation function derived in \cite{gmt09}. 
 
The Boltzmann-Langevin equation is the
starting point for the derivation of fluctuating hydrodynamic equations. This
can be done by projecting the Boltzmann-Langevin equation into the
hydrodynamic subspace. As our knowledge of the spectrum of the linearized
Boltzmann-Fokker-Planck operator is quite limited, we have focused on the
equation for the transverse velocity field,  that is decoupled from the rest of
the fluctuating hydrodynamic equations. This specific case was studied in
\cite{bgm08,bmg09} for the free cooling state, where it was shown
that the relevant Langevin noise 
is not white and that there is no a fluctuation-dissipation
relation. In other words, the amplitude of the noise is not related to the shear
viscosity. On the other hand, in the 
stochastically heated system, the behavior that we have reported is different. 
First, the noise of the transverse velocity contains two parts: one coming
from the thermostat (which does not conserve momentum locally) together 
with the more standard
fluctuating pressure tensor. 
The correlation function of the noise can then be written as a sum 
of direct and cross terms made up from the previous two contributions.
Moreover, in contrast to the free cooling scenario, the noise can be
considered as white \cite{rque}, as the
dynamics of the velocity is as slow as desired in the hydrodynamic (low
$k$) limit.  As in the free cooling case, in the hydrodynamic limit, the 
amplitude of the noise is {\em priori} not related to the shear
viscosity (such a relation, of fluctuation-dissipation type,
had been assumed in the approach of Ref \cite{vnetp99}). However, considering that the two-particle 
velocity correlation function has only hydrodynamic modes --which seems
a reasonable assumption-- somehow restores fluctuation-dissipation and
we obtain the
expression assumed in \cite{vnetp99}, with the actual inelastic shear
viscosity. In principle, this rather surprising result --
reminiscent of those reported in Refs \cite{Puglisi02}--
is not limited to small
inelasticity.

For future work, remains the extension of the theory to the other
hydrodynamic equations (beyond the transverse velocity), together with a 
generalization of the scheme presented here to a more
general class of thermostated systems, such as the ones with
multiplicative noise \cite{cafiero}.

\section{Acknowledgments}

This research was supported by the Ministerio de Educaci\'{o}n y
Ciencia (Spain) through Grant No. FIS2008-01339 (partially financed
by FEDER funds). We acknowledge financial support from Becas de
la Fundaci{\'o}n La Caixa and from Agence
Nationale de la Recherche (grant ANR-05-JCJC-44482).

\appendix

\section{Derivation of the transverse velocity field equation}\label{apendiceA}
In this Appendix we derive the equation for the transverse velocity
field, $w_\perp(\mathbf{k},s)$, in the hydrodynamic limit. The starting point
is equation (\ref{III.7})
\begin{eqnarray}\label{ec_v1}
\left[\frac{\partial}{\partial s}-P\Lambda(\mathbf{k},\mathbf{c})\right]
P\delta\tilde{F}(\mathbf{k},\mathbf{c},s)
&+&P(\mathbf{k}\cdot\mathbf{c})\int_0^s ds^\prime 
e^{Q\Lambda(\mathbf{k},\mathbf{c})(s-s^\prime)}
Q(\mathbf{k}\cdot\mathbf{c})P\delta\tilde{F}(\mathbf{k},\mathbf{c},s^\prime)
\nonumber\\
&=&P\tilde{R}(\mathbf{k},\mathbf{c},s)\nonumber\\
&-&Pi(\mathbf{k}\cdot\mathbf{c})\int_0^s ds^\prime 
e^{Q\Lambda(\mathbf{k},\mathbf{c})(s-s^\prime)}
Q\tilde{R}(\mathbf{k},\mathbf{c},s^\prime).
\end{eqnarray}

Let us first consider the term 
$P\Lambda(\mathbf{k},\mathbf{c})P\delta\tilde{F}(\mathbf{k},\mathbf{c},s)$. As
$\Lambda(\mathbf{c})\xi_{2\perp}(\mathbf{c})=0$ and  
$\int d\mathbf{c}(\mathbf{\hat{k}_\perp}\cdot\mathbf{c})
(\mathbf{\hat{k}}\cdot\mathbf{c})\xi_{2\perp}(\mathbf{c})=0$, we easily
have 
\be\label{B.1}
P\Lambda(\mathbf{k},\mathbf{c})P\delta\tilde{F}(\mathbf{k},\mathbf{c},s)=0. 
\ee

Let us evaluate the last term of the left hand side of Eq. (\ref{ec_v1}). 
To second order in $k$, we have
\begin{eqnarray}\label{B.3}
P(\mathbf{k}\cdot\mathbf{c})\int_0^s ds^{\prime}
e^{Q\Lambda(\mathbf{k},\mathbf{c})(s-s^\prime)}Q(\mathbf{k}\cdot\mathbf{c})
P\delta\tilde{F}(\mathbf{k},\mathbf{c},s^\prime)\nonumber\\
\simeq k^2\xi_{2\perp}(\mathbf{c})\int_0^sds'w_\perp(\mathbf{k},s')
\int d\mathbf{c}
(\mathbf{\hat{k}}\cdot\mathbf{c})(\mathbf{\hat{k}_\perp}\cdot\mathbf{c})
e^{\Lambda(\mathbf{c})(s-s')}\mathbf{\hat{k}}\cdot\mathbf{c}
\xi_{2\perp}(\mathbf{c})\nonumber\\
=k^2\xi_{2\perp}(\mathbf{c})\int_0^sds'w_\perp(\mathbf{k},s')G_{xy}(s-s')
\end{eqnarray}
where we have introduced
\begin{eqnarray}\label{B.4}
G_{xy}(s)\equiv\int d\mathbf{c}(\mathbf{\hat{k}_\perp}\cdot\mathbf{c})
(\mathbf{\hat{k}}\cdot\mathbf{c})e^{\Lambda(\mathbf{c})s}
(\mathbf{\hat{k}}\cdot\mathbf{c})\xi_{2\perp}(\mathbf{c})\nonumber\\
=\int d\mathbf{c}c_xc_ye^{\Lambda(\mathbf{c})s}c_x\xi_{2y}(\mathbf{c}),
\end{eqnarray}
and use has been made of the fact that the operator $\Lambda(\mathbf{c})$
is isotropic. In the hydrodynamic limit, the velocity evolves in a scale much
slower that the scale in which the function $G_{xy}(s)$ decays. We then have
\begin{equation}\label{B.5}
\int_0^s ds^\prime w_\perp(\mathbf{k},s^\prime)G_{xy}(s-s^\prime)
\to\tilde{\eta}w_\perp(\mathbf{k},s), 
\end{equation}
where $\tilde{\eta}$ is the dimensionless shear viscosity
\begin{equation}\label{B.6}
\tilde{\eta}=\int_0^\infty ds G_{xy}(s).
\end{equation}

The noise terms are the last two terms of Eq. (\ref{ec_v1})
\begin{eqnarray}\label{B.7}
P\tilde{R}(\mathbf{k},\mathbf{c},s)&=&\xi_{2\perp}(\mathbf{c})
\int d\mathbf{c}(\hat{\mathbf{k}}_\perp\cdot\mathbf{c})
\tilde{R}(\mathbf{k},\mathbf{c},s)\nonumber\\
&=&\xi_{2\perp}(\mathbf{c})S(\mathbf{k},s),
\end{eqnarray}
and 
\begin{eqnarray}\label{B.8}
&&P(i\mathbf{k}\cdot\mathbf{c})\int_0^s ds^\prime 
e^{Q\Lambda(\mathbf{k},\mathbf{c})(s-s^\prime)}
Q\tilde{R}(\mathbf{k},\mathbf{c},s^\prime)\nonumber\\
&&=\xi_{2\perp}(\mathbf{c}) ik\int d\mathbf{c}
(\hat{\mathbf{k}}_\perp\cdot\mathbf{c})
(\hat{\mathbf{k}}\cdot\mathbf{c})\int_0^s ds^\prime 
e^{Q\Lambda(\mathbf{k},\mathbf{c})(s-s^\prime)}
Q\tilde{R}(\mathbf{k},\mathbf{c},s^\prime)\nonumber\\
&&=-\xi_{2\perp}(\mathbf{c})\Pi(\mathbf{k},s), 
\end{eqnarray}
where we have used the definitions of $S(\mathbf{k},s)$ and the fluctuating
pressure tensor, $\Pi(\mathbf{k},s)$, Eqs. (\ref{III.10}) and (\ref{III.11}). 
Finally, by multiplying Eq. (\ref{ec_v1}) by 
$\mathbf{\hat{k}_\perp}\cdot\mathbf{c}$ and further integrating over
velocities, we obtain the equation of the transverse velocity of the main text.

\section{Autocorrelation function of $R_w(\mathbf{k},s)$}\label{apendiceB}
In this Appendix we evaluate the correlation function of the noise of the
transverse velocity field, $R_w(\mathbf{k},s)$. We consider
$\mathbf{k}\ne\mathbf{0}$, $s_1<s_2$ with $s_1$ large. 
It is convenient to introduce the following notation for the transverse and
parallel components of the vector $\mathbf{c}$
\be
\mathbf{\hat{k}_\perp}\cdot\mathbf{c}=c_\perp, \quad
\mathbf{\hat{k}}\cdot\mathbf{c}=c_\parallel. 
\ee
The autocorrelation function of $R_w(\mathbf{k},s)$ reads, in terms of
$S(\mathbf{k},s)$ and $\Pi(\mathbf{k},s)$, 
\begin{eqnarray}
  \label{eq:C.1}
  \langle R_w(\mathbf{k},s_1)R_w(-\mathbf{k},s_2)\rangle_H
&=&\langle(S(\mathbf{k},s_1)+\Pi(\mathbf{k},s_1))(S(-\mathbf{k},s_2)
+\Pi(-\mathbf{k},s_2))\rangle_H\nonumber\\
&=&\langle S(\mathbf{k},s_1)S(-\mathbf{k},s_2)\rangle_H
+\langle S(\mathbf{k},s_1)\Pi(-\mathbf{k},s_2)\rangle_H\nonumber\\
&+&\langle\Pi(\mathbf{k},s_1)S(-\mathbf{k},s_2)\rangle_H
+\langle \Pi(\mathbf{k},s_1)\Pi(-\mathbf{k},s_2)\rangle_H.\nonumber\\
\end{eqnarray}
We now calculate each correlation function taking into account the
microscopic expressions of $S(\mathbf{k},s)$ and $\Pi(\mathbf{k},s)$,
Eqs. (\ref{III.10}) and (\ref{III.11}), and the correlation function of the
noise of the Boltzmann-Langevin equation, Eqs. (\ref{II.17}) and
(\ref{II.18}). 

The first term is 
\begin{eqnarray}\label{eq:C.2}
\langle S(\mathbf{k},s_1)S(-\mathbf{k},s_2)\rangle_H&=&\int d\mathbf{c}_1
\int d\mathbf{c}_2c_{1\perp}c_{2\perp}
\langle\tilde{R}(\mathbf{k},\mathbf{c}_1,s_1)
\tilde{R}(-\mathbf{k},\mathbf{c}_2,s_2)\rangle_H\nonumber\\
&=&\frac{\tilde{V}^2}{N}\delta(s_1-s_2)\int d\mathbf{c}_1\int d\mathbf{c}_2
c_{1\perp}c_{2\perp}\Gamma(\mathbf{c}_1,\mathbf{c}_2)\nonumber\\
&=&\frac{\tilde{V}^2}{N}\delta(s_1-s_2)\int d\mathbf{c}_1\int d\mathbf{c}_2 
c_{1\perp}c_{2\perp}\bar{T}(\mathbf{c}_1,\mathbf{c}_2)
\chi_H(c_1)\chi_H(c_2)\nonumber\\
&=&\tilde{\xi}^2\frac{\tilde{V}^2}{N}\delta(s_1-s_2).
\end{eqnarray}
where we have used the relation 
$\int d\mathbf{c}_1\int d\mathbf{c}_2c_{1\perp}c_{2\perp}
\bar{T}(\mathbf{c}_1,\mathbf{c}_2)\chi_H(c_1)\chi_H(c_2)=\tilde{\xi}^2$, that
is proved in \cite{gmt09}. 

The second correlation function is
\begin{eqnarray}\label{C.10}
\langle S(\mathbf{k},s_1)\Pi(-\mathbf{k},s_2)\rangle_H
\qquad\qquad\qquad\qquad\qquad\qquad\qquad\qquad\qquad\qquad\qquad\qquad
\qquad\qquad
\nonumber\\
=\int d\mathbf{c}_1c_{1\perp}
 ik\int_0^{s_2}ds
\int d\mathbf{c}_2c_{2\parallel}c_{2\perp}
e^{Q_2\Lambda(-\mathbf{k},\mathbf{c}_2)(s_2-s)}\langle
\tilde{R}(\mathbf{k},\mathbf{c}_1,s_1)Q_2
\tilde{R}(-\mathbf{k},\mathbf{c}_2,s)\rangle_H\nonumber\\
=ik\frac{\tilde{V}^2}{N}\int d\mathbf{c}_1\int d\mathbf{c}_2
c_{1\perp}c_{2\parallel}c_{2\perp}
e^{Q_2\Lambda(-\mathbf{k},\mathbf{c}_2)
(s_2-s_1)}Q_2\Gamma(\mathbf{c}_1,\mathbf{c}_2).
\end{eqnarray}
Here, we have changed the sign of $\Pi(-\mathbf{k},s)$ because we are
dealing with the $-\mathbf{k}$ component and then $\mathbf{\hat{k}}\to
-\mathbf{\hat{k}}$ (we do not change 
$\mathbf{\hat{k}}_\perp\to -\mathbf{\hat{k}}_\perp$, because this vector comes
from the projector $P$ and it is fixed). 

The third term vanishes
\begin{eqnarray}\label{eq:C.3}  
\langle\Pi(\mathbf{k},s_1)S(-\mathbf{k},s_2)\rangle_H
&=&-ik\int_o^{s_1}ds^{\prime}\int d\mathbf{c}_1c_{1\parallel}c_{1\perp}
\int d\mathbf{c}_2c_{2\perp}\nonumber\\
&\times&e^{Q_1\Lambda(\mathbf{k},\mathbf{c}_1)(s-s^\prime)}
Q_1\langle\tilde{R}(\mathbf{k},\mathbf{c}_1,s^\prime)
\tilde{R}(-\mathbf{k},\mathbf{c}_2,s_2)\rangle_H=0,\nonumber\\
\end{eqnarray}
because $\langle\tilde{R}(s^\prime)\tilde{R}(s_2)\rangle_H=0$ for 
$s^\prime\in(0,s_1)$ with $s_1<s_2$.

Finally, we evaluate the last term to second order in $k$
\begin{eqnarray}\label{C.4}
&&\langle \Pi(\mathbf{k},s_1)\Pi(-\mathbf{k},s_2)\rangle_H
\nonumber\\
&&\simeq k^2\int_0^{s_1}ds_1^\prime \int_0^{s_2}ds_2^\prime\int d\mathbf{c}_1
\int d\mathbf{c}_2 c_{1\parallel}c_{1\perp}c_{2\parallel} c_{2\perp}
e^{\Lambda(\mathbf{c}_1)(s_1-s_1^\prime)+\Lambda(\mathbf{c}_2)(s_2-s_2^\prime)}
\nonumber\\
&&\times\langle\tilde{R}(\mathbf{k},\mathbf{c}_1,s_1^\prime)
\tilde{R}(-\mathbf{k},\mathbf{c}_2,s_2^\prime)\rangle_H\nonumber\\
&&=\frac{\tilde{V}^2}{N}k^2\int_0^{s_1}ds_1^\prime \int_0^{s_2}ds_2^\prime
\int d\mathbf{c}_1\int d\mathbf{c}_2c_{1\parallel}c_{1\perp}c_{2\parallel} 
c_{2\perp}e^{\Lambda(\mathbf{c}_1)(s_1-s_1^\prime)
+\Lambda(\mathbf{c}_2)(s_2-s_2^\prime)}\Gamma(\mathbf{c}_1,\mathbf{c}_2)
\delta(s_1^\prime-s_2^\prime)\nonumber\\
&&=\frac{\tilde{V}^2}{N}k^2\int d\mathbf{c}_1\int d\mathbf{c}_2
c_{1\parallel}c_{1\perp}c_{2\parallel} c_{2\perp}\int_0^{s_1}ds 
e^{\Lambda(\mathbf{c}_1)(s_1-s)+\Lambda(\mathbf{c}_2)(s_2-s)}
\Gamma(\mathbf{c}_1,\mathbf{c}_2)\nonumber\\
&&=\frac{\tilde{V}^2}{N}k^2\int d\mathbf{c}_1\int d\mathbf{c}_2c_{1\parallel}
c_{1\perp}c_{2\parallel} c_{2\perp}e^{\Lambda(\mathbf{c}_1)s_1
+\Lambda(\mathbf{c}_2)s_2}\int_0^{s_1}dse^{-s[\Lambda(\mathbf{c}_1)
+\Lambda(\mathbf{c}_2)]}\Gamma(\mathbf{c}_1,\mathbf{c}_2)\nonumber\\
&&\simeq\frac{\tilde{V}^2}{N}k^2\int d\mathbf{c}_1\int 
d\mathbf{c}_2c_{1\parallel}c_{1\perp}c_{2\parallel} c_{2\perp}
e^{\Lambda(\mathbf{c}_2)(s_2-s_1)}\nonumber\\
&&\times[\Lambda(\mathbf{c}_1)
+\Lambda(\mathbf{c}_2)]^{-1}
\left[-\Gamma(\mathbf{c}_1,\mathbf{c}_2)+\tilde{\xi}^2
\frac{\partial}{\partial \mathbf{c}_1}\cdot
\frac{\partial}{\partial\mathbf{c}_2}
\chi_H(c_1)\chi_H(c_2)\right], 
\end{eqnarray}
where, in the last step, we have taken into account that $s_1$ is large and
we have introduced the term 
$\frac{\partial}{\partial\mathbf{c}_1}\cdot
\frac{\partial}{\partial\mathbf{c}_2}\chi_H(c_1)\chi_H(c_2)$. This term does
not contribute to the integral, but is written for convenience,
to make a connection with the global correlation 
function $\phi_H(\mathbf{c}_1,\mathbf{c}_2)
\equiv\int d\mathbf{l}h_H(\mathbf{l},\mathbf{c}_1,\mathbf{c}_2)$ 
which fulfills Eq. (\ref{ec_phi}). 
In doing so, we find that the autocorrelation function of $\Pi(\mathbf{k},s)$
reads
\begin{equation}\label{C.8}
\langle \Pi(\mathbf{k},s_1)\Pi(-\mathbf{k},s_2)\rangle_H\simeq 
\frac{\tilde{V}^2}{N}k^2C_{xy}(s_2-s_1),
\end{equation}
where 
\begin{equation}\label{C.9}
C_{xy}(s_2-s_1)=\int d\mathbf{c}_1\int d\mathbf{c}_2 
c_{1x}c_{1y}c_{2x}c_{2y}e^{\Lambda_2(s_2-s_1)}
\phi_H(\mathbf{c}_1,\mathbf{c}_2). 
\end{equation}
As $\phi_H(\mathbf{c}_1,\mathbf{c}_2)$ is the integral of the correlation
function $h_H(\mathbf{l},\mathbf{c}_1,\mathbf{c}_2)$, $C_{xy}$ can be
identified as the correlation function of the global quantity 
$\sum_{i=1}^NV_x(t)V_{y}(t)$.

\section{}\label{apendiceC}
In this Appendix we evaluate the time integral of the correlation function
$\langle S(\mathbf{k},0)\Pi(-\mathbf{k},s)\rangle_H$ in the hydrodynamic
limit. Using the notation of the previous Appendix, we have
\begin{eqnarray}\label{D.1}
&&\int_0^\infty ds\langle S(\mathbf{k},0)\Pi(-\mathbf{k},s)\rangle_H\nonumber\\
&&=ik\frac{\tilde{V}^2}{N}\int d\mathbf{c}_1\int d\mathbf{c}_2
c_{1\perp}c_{2\parallel}c_{2\perp}
\left[\frac{e^{Q_2\Lambda(-\mathbf{k},\mathbf{c}_2)s}}
{Q_2\Lambda(-\mathbf{k},\mathbf{c}_2)}\right]_0^\infty
Q_2\Gamma(\mathbf{c}_1,\mathbf{c}_2)\nonumber\\   
&&=-ik\frac{\tilde{V}^2}{N}\int d\mathbf{c}_1\int d\mathbf{c}_2 
c_{1\perp}c_{2\parallel}c_{2\perp}
\frac{1}{Q_2\Lambda(-\mathbf{k},\mathbf{c}_2)}Q_2
\Gamma(\mathbf{c}_1,\mathbf{c}_2)\nonumber\\
&&=ik\frac{\tilde{V}^2}{N}\int d\mathbf{c}_1\int d\mathbf{c}_2 
c_{1\perp}c_{2\parallel}c_{2\perp}
\frac{Q_2[\Lambda(\mathbf{k},\mathbf{c}_1)+\Lambda(-\mathbf{k},\mathbf{c}_2)]}
{Q_2\Lambda(-\mathbf{k},\mathbf{c}_2)}
\phi_H(\mathbf{k},\mathbf{c}_1,\mathbf{c}_2),\nonumber\\
\end{eqnarray}
where we have introduced the function 
\be
\phi_H(\mathbf{k},\mathbf{c}_1,\mathbf{c}_2)
=\int d\mathbf{l}e^{-i\mathbf{k}\cdot\mathbf{l}}
h_H(\mathbf{l},\mathbf{c}_1,\mathbf{c}_2), 
\ee 
that fulfills the Fourier transform of Eq. (\ref{II.9})
\be
[\Lambda(\mathbf{k},\mathbf{c}_1)+\Lambda(-\mathbf{k},\mathbf{c}_2)]
\phi_H(\mathbf{k},\mathbf{c}_1,\mathbf{c}_2)=
-\Gamma(\mathbf{c}_1,\mathbf{c}_2)+\tilde{\xi}^2
\frac{\partial}{\partial \mathbf{c}_1}\cdot
\frac{\partial}{\partial\mathbf{c}_2}\chi_H(c_1)\chi_H(c_2)\delta(\mathbf{k}). 
\ee
Note that the last term in the previous equation only appears for
$\mathbf{k}=\mathbf{0}$. 
Taking into account that $c_\perp\chi_H(c)$ is left eigenfunction associated
to the null eigenvalue and after some algebra we obtain
\begin{eqnarray}\label{D1b}
&&\int_0^\infty ds\langle S(\mathbf{k},0)\Pi(-\mathbf{k},s)\rangle_H\nonumber\\
&&=\frac{\tilde{V}^2}{N}k^2\int d\mathbf{c}_1\int d\mathbf{c}_2 
c_{1\parallel}c_{1\perp}c_{2\parallel}c_{2\perp}
\frac{1}{Q_2\Lambda(-\mathbf{k},\mathbf{c}_2)}
\phi_H(\mathbf{k},\mathbf{c}_1,\mathbf{c}_2)
\nonumber\\
&&+\frac{\tilde{V}^2}{N}ik\int d\mathbf{c}_1\int d\mathbf{c}_2
c_{1\perp}c_{2\parallel}c_{2\perp}
\phi_H(\mathbf{k},\mathbf{c}_1,\mathbf{c}_2).
\end{eqnarray}
Now, let us consider the hydrodynamic limit of (\ref{D1b}). The first term
gives 
\begin{equation}\label{D.2}
\frac{\tilde{V}^2}{N}k^2\int d\mathbf{c}_1\int d\mathbf{c}_2 
c_{1\parallel}c_{1\perp}c_{2\parallel}c_{2\perp}
\frac{1}{Q_2\Lambda(-\mathbf{k},\mathbf{c}_2)}
\phi_H(\mathbf{k},\mathbf{c}_1,\mathbf{c}_2)
\to-\frac{\tilde{V}^2}{N}k^2\int_0^\infty dsC_{xy}(s). 
\end{equation}
The second term can be evaluated by using the following expansion in powers of
$k$ 
\be
\left[\Lambda(\mathbf{c})-i\mathbf{k}\cdot\mathbf{c}\right]^{-1}\simeq
\Lambda(\mathbf{c})^{-1}
+\Lambda(\mathbf{c})^{-1}(i\mathbf{k}\cdot\mathbf{c})\Lambda(\mathbf{c})^{-1}, 
\ee
which yields
\begin{eqnarray}\label{D.4}
\frac{\tilde{V}^2}{N}ik\int d\mathbf{c}_1\int d\mathbf{c}_2
c_{1\perp}c_{2\parallel}c_{2\perp}
\phi_H(\mathbf{k},\mathbf{c}_1,\mathbf{c}_2)\nonumber\\
\to\frac{\tilde{V}^2}{N}k^2\int_0^\infty dsC_{xy}(s)
-\frac{\tilde{V}^2}{N}k^2\int d\mathbf{c}_1\int d\mathbf{c}_2
c_{1x}c_{2x}c_{2y}\int_0^\infty ds e^{\Lambda(\mathbf{c}_2)s}c_{2y}
\phi_H(\mathbf{c}_1,\mathbf{c}_2). \nonumber \\
\end{eqnarray}
Taking into account (\ref{D.2}) and (\ref{D.4}), we obtain
\begin{eqnarray}\label{D.5}
\int_0^\infty ds\langle S(\mathbf{k},0)\Pi(-\mathbf{k},s)\rangle
&\to&-\frac{\tilde{V}^2}{N}k^2\int d\mathbf{c}_1\int d\mathbf{c}_2
c_{1x}c_{2x}c_{2y}\int_0^\infty ds e^{\Lambda(\mathbf{c}_2)s}c_{2y}
\phi_H(\mathbf{c}_1,\mathbf{c}_2)\nonumber\\
&=&\frac{\tilde{V}^2}{N}k^2\int d\mathbf{c}_1\int d\mathbf{c}_2
c_{1x}c_{2x}c_{2y}\Lambda(\mathbf{c}_2)^{-1}c_{2y}
\phi_H(\mathbf{c}_1,\mathbf{c}_2).
\end{eqnarray}

\section{}\label{apendiceD}
In this Appendix we evaluate the $k^2$ component of the correlation function
of $R_w$ in terms of the one and two-particle distribution functions, $\chi_H$
and $\chi_2$
\be
\phi_H(\mathbf{c}_1,\mathbf{c}_2)
=\chi_H(c_1)\delta(\mathbf{c}_{12})+\chi_2(\mathbf{c}_1,\mathbf{c}_2). 
\ee
The first term is
\begin{eqnarray}\label{E.1}
\int_0^\infty dsC_{xy}(s)&=&\int_0^\infty ds\int d\mathbf{c}_1\int 
d\mathbf{c}_2
c_{1x}c_{1y}c_{2x}c_{2y}
e^{s\Lambda(\mathbf{c}_2)}\phi_H(\mathbf{c}_1,\mathbf{c}_2)\nonumber\\
&=&-\int d\mathbf{c}_1\int d\mathbf{c}_2 
c_{1x}c_{1y}c_{2x}c_{2y}
\Lambda(\mathbf{c}_2)^{-1}\phi_H(\mathbf{c}_1,\mathbf{c}_2)\nonumber\\
&=& -\int d\mathbf{c}c_xc_y\Lambda(\mathbf{c})^{-1}c_xc_y\chi_H(c)\nonumber\\
&-&\int d\mathbf{c}_1\int d\mathbf{c}_2 c_{1x}c_{1y}c_{2x}c_{2y}
\Lambda(\mathbf{c}_2)^{-1}\chi_2(\mathbf{c}_1,\mathbf{c}_2). 
\end{eqnarray}
If we do the same in the second term, we have
\begin{eqnarray}
\int d\mathbf{c}_1\int d\mathbf{c}_2 c_{1x}c_{2x}c_{2y}
\Lambda(\mathbf{c}_2)^{-1}c_{2y}\phi_H(\mathbf{c}_1,\mathbf{c}_2)
&=&\int d\mathbf{c}c_xc_y\Lambda(\mathbf{c})^{-1}c_xc_y\chi_H(c)\nonumber\\
&+&\int d\mathbf{c}_1\int d\mathbf{c}_2 c_{1x}c_{2x}c_{2y}
\Lambda(\mathbf{c}_2)^{-1}c_{2y}\chi_2(\mathbf{c}_1,\mathbf{c}_2).\nonumber\\
\end{eqnarray}
It can be seen that the sum of the two terms only depends on the two-particle
correlation function and we obtain the $k^2$ part of Eq. (\ref{III.16}).


\begin{thebibliography}{10}
\bibitem{rque}
Note that at the Boltzmann-Langevin level, the noise
is assumed white ; in the present driven case, it is then shown that
the noise for the fluctuating transverse velocity can be considered
as white also ; this was not the case in the free cooling situation 
\cite{bmg09}.
\end{thebibliography}
\end{document}